\documentclass[aps,prl,twocolumn,groupedaddress]{revtex4}
\usepackage{graphicx}

\begin{document}
\title{Network rigidity and dynamics of oxides}
\author{Kostya Trachenko and Martin T Dove}
\address{Department of Earth Sciences, University of Cambridge,
Downing Street, Cambridge, CB2~3EQ, UK}

\begin{abstract}
If a hierarchy of interatomic interactions exists in a solid, low-frequency modes can be found from viewing this solid as a mechanical network. In this case, the low-frequency modes are determined by the network rigidity. We study the low-frequency modes (rigid unit modes, or RUMs) in several important oxide materials and discuss how the RUMs affect their properties. In SiO$_2$ glass, the ability to support RUMs governs its relaxation in the wide range of pressures and temperatures, giving rise to the non-trivial pressure window. It also affects other properties, including crystallization, slow relaxation and compressibility. At ambient pressure and low temperature, RUM flexibility is related to the large-scale localized atomic motions. Whether these motions are interpreted as independent two-level systems or collective density excitations, the RUM flexibility determines whether and to what extent low-energy excitations can exist in a given glass structure. Finally, we discuss the RUM in, perhaps unexpectedly, cuprate superconductors, and its relevance for superconductivity, including the d-wave symmetry of the order parameter and other properties.
\end{abstract}

\maketitle

\section{Introduction and overview}

If the energy to stretch a chemical bond between two atoms considerably exceeds the thermal energy, such a bond can be viewed
as a Lagrangian constraint, in a sense that it keeps two atoms at a fixed distance. This idea can be made useful in the case of covalent materials, in which two-body stretching and three-body bending forces considerably exceed all others. These short-range
interactions can be translated into the building blocks of a mechanical network. This has been the starting point of the Phillips theory of network glasses \cite{phillips}. By requiring that the number of degrees of freedom is equal to the average number of
bonding constraints, this theory predicted the average coordination number $\langle r\rangle$ for which glass forming ability is optimized. Since then, this picture has been widely used to discuss relaxation in covalent glasses and crystals.

The constraint theory offers a great reduction in treating interactions in a solid, by translating them into a mechanical network. The building blocks of such a network, two- and three-body elements, represent strong two-body stretching and three-body bending interactions, and all remaining weaker forces (see below) are ignored. One method of treating excitations in such a network is known as Maxwell counting \cite{max}, which makes predictions about the network low-energy states. Any modes in a mechanical network that keep local constraints intact, have zero frequency because there is no restoring forces to such deformations. According to Maxwell counting, the number of such modes is equal to the difference between the number of degrees of freedom, $N_f$, and the number of bonding constraints, $N_c$. Therefore, the existence of the hierarchy of interactions in a solid can have important implications for the hierarchy of vibrational modes in terms of their frequency. In the simulation study of constraint counting in glasses, ``floppy'' modes appear when the network becomes under-constrained, $N_c < N_f$, or $\langle r\rangle <2.4$ \cite{th0,thor-coll} (the term ``floppy'' here points to the fact that in real systems, weaker interactions, e.g., Van-Der-Waals forces, always give a non-zero restoring force associated with propagation of constraint-obeying modes, making their frequency not zero exactly, but some small values).

By construction, the picture which maps interatomic interactions into a network of mechanical constraints, is limited to solids with short-range covalent interactions. If ionic contribution to bonding is substantial, the mapping of interatomic interactions into a network is problematic due to the long-range nature of Coloumb forces and the absence of the hierarchy of interactions
\cite{thor-coll}. It is nevertheless still possible to consider many important properties of solids with substantial ionic contribution to bonding, using a general idea that a certain chemical interaction can be mapped into a mechanical constraint. Consider very common silica glass. Ionic contribution to Si-O bond is at least as strong as covalent one, resulting in the fact that although O-Si-O bending constraint is intact, Si-O-Si angular constraint is broken, as is seen by the very broad distribution of Si-O-Si angles \cite{mozzi}. Hence the usual constraint counting procedure would overestimate rigidity of silica glass. However, despite the substantial ionicity of Si-O bond, it is known from both experiments and computer simulations that SiO$_4$ tetrahedra are very rigid. This is related to the high energy cost involved in the deformation of the electronic density that has a tetrahedral symmetry. Hence if we are interested in low-energy vibrations of silica, its Phillips network analogue is a collection of SiO$_4$ rigid units, loosely connected at corners. Thus even though there is a considerable ionic contribution to bonding in a solid, the knowledge of its structure and chemistry can still allow us to map interatomic interactions into a generalized network, albeit with different building blocks: these do not correspond to two- and three-body constraints as in the
Phillips theory, but to local rigid units. The modes that propagate without the constituent units having to distort have been named Rigid Unit Modes (RUMs) \cite{r1,r11,r2,r3,r4}.

An example of a localized RUM in SiO$_2$ glass is shown in Figure 1. It shows the motion of a local cluster of connected rigid SiO$_4$ tetrahedra. It can be seen that atomic displacements are largest in the centre of the cluster, and decay at the cluster periphery. Another example of a RUM in a two dimensional crystalline system is shown in Figure 2. Such a mode exists in CuO$_2$ plane in cuprate superconductors. Unlike in Figure 1, all atoms in the structure participate equally in the RUM motion.

\begin{figure}
\begin{center}
{\scalebox{0.4}{\includegraphics{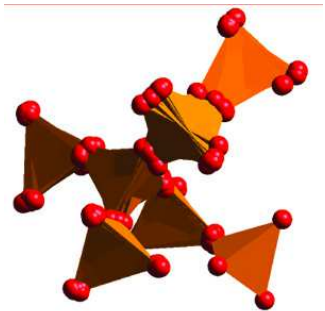}}}
\end{center}
\caption{A RUM in SiO$_2$ glass, showing local reorientations of rigid tetrahedra. A series of configurations through the RUM motion are shown.}
\label{fig1}
\end{figure}

\begin{figure}
\begin{center}
{\scalebox{0.53}{\includegraphics{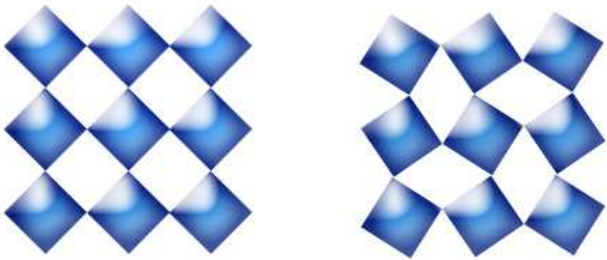}}}
\end{center}
\caption{An extended RUM in a two-dimensional system. In the case of CuO$_2$ plane in cuprate superconductors, Cu atoms are in the centre of a square, and O atoms are at the vertices.}
\label{fig2}
\end{figure}

There are many ways in which RUMs can affect important physical properties of a material. For example, RUMs have been shown to act as soft modes in structural phase transitions in many important minerals \cite{r1,r11}. As a result, the flexibility against RUMs has explained why some minerals are more prone to phase transition than others \cite{r1,r11}. Because RUMs give rise to strong dynamic disorder (see below), they have been used to interpret experiments aimed at elucidation of crystal structures. Here, RUM flexibility enables one to disentangle static structure from the effects of dynamic disorder \cite{r2}. RUM flexibility can also explain the ease with which framework minerals can incorporate different ions \cite{r3}. Finally, RUMs can explain an interesting effect of negative thermal expansion \cite{r4}. As a result of certain RUM deformations system dimensions decrease (see, for example, Figure 2). This effect increases with temperature, and if it is larger than usual thermal expansion, the net effect is the negative thermal expansion \cite{r4}.

The RUM-related effects discussed above have been studied in crystals. At the same time, the original constraint theory of Phillips-Thorpe was developed for glasses. It is particularly interesting to study glasses because this is an under-researched area in condensed matter physics. Indeed, solid state physics has traditionally discussed crystals, whereas topologically disordered solids have been barely mentioned in the most of the textbooks \cite{zallen}. Experimental data acquired in the last few decades have shown that topologically disordered matter and glasses in particular display new exciting effects not seen in crystals. A large body of experimental evidence concerning these effects has accumulated in the literature, but theoretical understanding of this data is lacking. This, in part, has stimulated our recent work on glasses.

In this contribution, we review our recent work on RUMs in glasses, with particular emphasis on their relationship with glass dynamics and relaxation behaviour. RUM flexibility has a profound effect on the dynamics, which we specify here from the outset. Because, by definition, RUMs involve only weak restoring forces, a RUM-flexible structure has large amplitudes of atomic displacements. This can also be seen from the well-known expression from the normal-mode analysis:

\begin{equation}
\langle Q_i^2\rangle\propto \frac{T}{\omega_i^2}
\end{equation}

\noindent where $Q_i$ is the normal mode coordinate. Because RUM flexibility implies the presence of the set of the rigid-unit normal modes with small frequencies $\omega_i$, it results in large $\langle Q_i^2\rangle$ and, therefore, large value of the atomic displacements.

We note that according to Eq. (1), the acoustic phonons with $\omega\sim 0$ trivially give large atomic displacements for any structure. By definition, these modes are RUMs because they do not involve distortion of the constituent units. However, as we will see below, RUM flexibility at ${\bf k}\sim 0$ is not related to relaxation phenomena of interest. On the other hand, interesting effects are often governed by RUM flexibility away from ${\bf k}\sim 0$, and it is this flexibility that differentiates one structure from another one, or affects structure properties under a varying physical parameter.

The outline of this paper is as follows. We begin the discussion with the ability of SiO$_2$ glass to support RUMs, and study how this ability is affected by pressure. The relaxation behaviour of SiO$_2$ glass under high pressure and temperature gives rise to a non-trivial pressure window. This window was predicted in our molecular dynamics simulations, and later confirmed experimentally. We explain the existence of the pressure window on the basis of the coupling of local relaxation events to the structural rigidity of glass. In this picture, the pressure window is centered at the rigidity percolation point, at which there is large reduction of RUMs in the densified glass structure. We also discuss the implications of RUM flexibility for other important properties as crystallization, slow relaxation and compressibility. The RUM flexibility is also related to the existence of two-level systems in SiO$_2$ glass, and we briefly discuss how these systems operate at the microscopic level. Finally, we discuss the RUM in, perhaps unexpectedly, cuprate superconductors, and its relevance for superconductivity.

\section{How to quantify the RUM flexibility}

For a given structure, RUMs can be found as low-frequency normal modes, using the standard lattice-dynamical programs. However, the proportion of these modes in the total density of states is very small. Consequently, if we want to study in some detail how RUMs change from one structure to another, or within the same structure under some modification (e. g., pressure), standard normal-mode analysis may not be suitable because the associated changes in the frequency spectrum would be small or even not noticeable. At the same time, changes of the RUM flexibility may be very important for dynamical and relaxation behaviour, for the reasons discussed in the previous section and below.

To overcome this problem and to facilitate the identification of RUMs, a special RUM detection tool has been developed. It is based on the ``split-atom'' method \cite{r1}, and has been implemented within the formalism of harmonic molecular lattice dynamics, using a program called CRUSH \cite{crush}. Noting that a RUM is a phonon normal mode that can propagate without requiring the distortions of the constituent rigid units, the task is to set up the dynamical matrix in such a way that RUMs are obtained as zero-frequency solutions. The constituent units (e.g., SiO$_4$ tetrahedra in silica or BO$_6$ octahedra in perovskite) are treated as rigid units within the framework of molecular lattice dynamics. The bridging oxygens are replaced by pairs of atoms that are associated with one tetrahedron or the other. The pairs of split atoms are held together by harmonic spring forces of zero equilibrium length, which act to resist any motion that moves them apart. The value of the spring constant sets the energy scale in the problem. Its value is usually fixed to give the maximal RUM frequency of about 1 THz, the typical frequency of low-energy vibrations measured experimentally \cite{ourjpcm}. The idea of the ``split-atom'' method is illustrated in Figure 3.

\begin{figure}
\begin{center}
{\scalebox{0.8}{\includegraphics{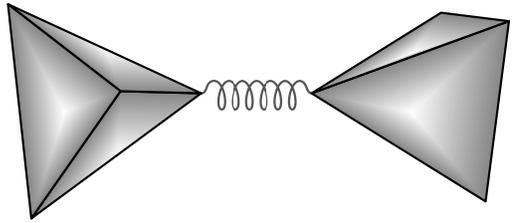}}}
\end{center}
\caption{Representation of the split-atom method. The spring has an equilibrium length of zero, and a force constant set to a value that best mimics the stiffness of the SiO$_4$ tetrahedra as judged from measurements of phonon frequencies.}
\label{fig3}
\end{figure}

In this method, RUMs are found as normal modes with exactly zero frequency. One way to quantify the presence of RUMs is through their number, by counting ${\bf k}$ points at which RUMs exist. Depending on the degree of flexibility, RUMs can be found for discrete sets of wave vectors \cite{r3}, lines of ${\bf k}$ points \cite{r6} or, as more recently found, for exotic two-dimensional surfaces in reciprocal space \cite{r4}. Another way to quantify the presence of RUMs is to construct their density of states $g(\omega)$ using the grid of random ${\bf k}$ points. In this case, the RUM flexibility is quantified by the behaviour of $g(\omega)$ at $\omega=0$: generally, $g(\omega)=const$ and $g(\omega)=0$ at $\omega=0$ correspond to the structure being flexible and non-flexible against RUMs, respectively. There are also intermediate cases that correspond to different levels of RUM flexibility that correspond to $g(\omega)\propto\omega$ and $g(\omega)\propto\omega^2$, discussed below in more detail.

It should be noted that the $g(\omega)\propto\omega$ produced this way is not to be confused with the $g(\omega)\propto\omega$ of a real material, because in a real material the full set of force constants would give nonzero values of the RUM frequencies and so we would not get $g(\omega)=const$ for the RUM-flexible structure. Instead, the RUM $g(\omega)$ should be seen as a particular diagnostic tool with the purpose of giving a unique quantitative assessment of the RUM flexibility of a structure.

\section{Network flexibility of SiO$_2$ glass}

Silica is perhaps the most important glass from the technological points of view. This glass has received the largest attention in both theoretical and experimental studies in physics, chemistry, materials science, mineralogy, engineering and other disciplines. In the next sections, we discuss how RUM flexibility affects the properties of SiO$_2$ glass.

\subsection{Maxwell counting and network flexibility}

Whether an infinite framework of corner-linked SiO$_4$ tetrahedra can vibrate without the tetrahedra distorting is actually a very subtle issue. The number of zero-frequency modes is equal to the difference between the number of degrees of freedom
and the number of constraints, a procedure known as ``Maxwell counting'' \cite{maxw}. The Maxwell counting procedure gives an interesting result when applied to silica: for a network of vertex-connected tetrahedra the number of degrees of freedom is equal to the number of constraints. For a SiO$_4$ tetrahedron, the integrity of the tetrahedron is fully defined by the fixing the lengths of the six O–-O distances and three of the Si-–O bonds, thus giving nine constraints. The number of degrees of freedom of a tetrahedron in the SiO$_2$ network is nine (accounting for the fact that a bridging O atom gives 3/2 of its degrees of freedom to each tetrahedron it is connected to), thus for silica there are nine constraints and nine degrees of freedom associated with each SiO$_2$. The same result can be obtained by considering the SiO$_4$ tetrahedra as rigid units with six (translational and rotational) degrees of freedom. Holding the corner of any tetrahedron at the same position as the corner of its neighbouring tetrahedron gives three constraints per corner, and noting that each corner is shared by two tetrahedra, the number of constraints is six (4$\times$3/2), or the same as the number of degrees of freedom. Either analysis assumes an infinite system, since the non-bridging bonds at surfaces will reduce the number of constraints per atom. By considering bonding constraints of individual tetrahedra, this analysis also accounts for the fact that Si-O-Si bending constraints are broken in SiO$_2$ glass.

In crystalline silicates it has been conjectured that symmetry may be responsible for degenerate constraints \cite{r1}. This may
lead to the increased ability of the structure to sustain floppy modes. In glass, there may also be degenerate constraints leading to the appearance of floppy modes. However, because of the topological disorder, the speculation about the possible reduction of the number of constraints is not straightforward. Thus one cannot easily predict whether the floppy modes can exist in silica glass. This is where the method described in the previous section comes useful.

\subsection{RUM flexibility}

The starting configurations of SiO$_2$ glass were obtained using classical molecular dynamics simulations. For details of structure generation, empirical potentials and other simulation details, see Refs \cite{ourjpcm,prl98}.

In Figure 4, we plot the RUM density of states calculated for $\beta$-cristobalite and SiO$_2$ glass \cite{ourjpcm,prl98}. The similarity of $g(\omega)$ for the two systems for $\omega=0$ is striking - in fact, one can view the overall form of $g(\omega)$ of silica glass simply as a lower-resolution version of $g(\omega)$ of $\beta$-cristobalite. This comparison implies that silica glass has the same RUM flexibility as $\beta$-cristobalite, which is actually an astonishing result given that the RUM flexibility of $\beta$-cristobalite had previously been interpreted as being due to the effects of the high symmetry of its crystal structure \cite{r1,r2,r7}.

\begin{figure}
\begin{center}
\rotatebox{-90}{\scalebox{0.4}{\includegraphics{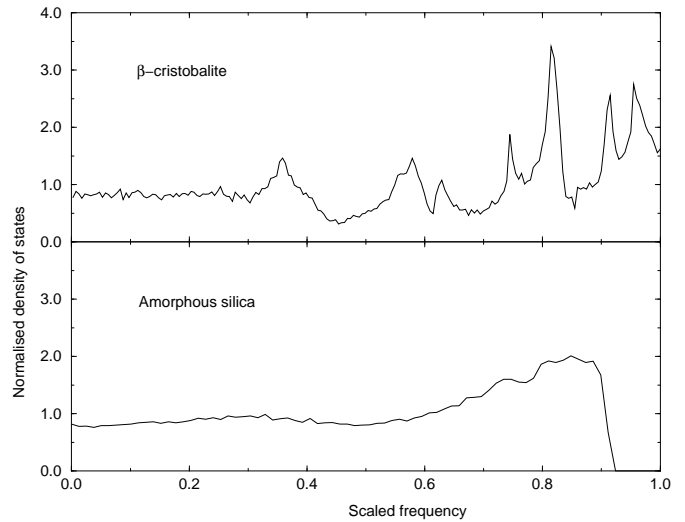}}}
\end{center}
\caption{The RUM density of states, calculated for $\beta$-cristobalite and silica glass. The values of the frequencies are determined by the spring force constant in the split-atom method, and have been scaled against the maximum frequency in this figure.}
\label{fig4}
\end{figure}

Experimentally, the conclusion about significant RUM flexibility is confirmed by our neutron scattering experiments, showing the presence of strong scattering in the low 0--5 meV region \cite{ourjpcm}.

The calculation of the participation ratio shows that that the low-$\omega$ RUM-like vibrations are not localized to any particular flexible segments of the glass structure, but involve mostly all tetrahedra \cite{ourjpcm,prl98}. This is an important insight which we will use in the discussion below.

\section{Network flexibility of SiO$_2$ glass under pressure}

\subsection{Why study network flexibility under pressure?}

The effects of pressure on crystals have been studied in detail and are well-known. On the other hand, the pressure response of amorphous solids is not understood well \cite{bra-jpcm}. Some of the new, as compared to crystals, features of high-pressure behaviour in amorphous solids include gradual coordination changes, long tails of transformations, slow logarithmic relaxation, permanent densification on pressure release, and others. New terms have been introduced in the area to describe these effects, such as ```polyamorphism'', `amorphous-amorphous transformation'' and ``low- and high-density amorphs'', and the possibility of the phase transitions between them has been studied and discussed \cite{bra-jpcm}. Generally, it is interesting to understand how the presence of topological disorder gives rise to new effects under pressure. This interest has stimulated our work aimed at better understanding of pressure-induced transformations in amorphous solids. In the following sections, we will discuss how RUM flexibility governs relaxation under high pressure and temperature.

Non-equilibrium relaxation phenomena in glasses are also related to the problem of glass transition, the most important and controversial problems in condensed matter physics \cite{phil1,langer}. It is believed that only high-temperature superconductivity can compete with glass transition in terms of controversy and involvement, although glass transition is a much older problem: the first anomalous ``glassy'' relaxation laws were established more than 150 years ago \cite{phil1}. Here, the origin of slow relaxation in supercooled liquids is the central open question \cite{phil1,langer}. Because it is also observed in glasses \cite{phil1}, it is believed that the common physical mechanism that sets the slow relaxation operates in both liquids and glasses. In the sections below, we will discuss how RUM flexibility is related to slow relaxation.

\subsection{RUM flexibility under pressure}

Using molecular dynamics simulations, we have compressed SiO$_2$ structure to different pressures. A detailed analysis of the compressed structure reveals that tetrahedral topology of glass does not change up to 3 GPa. The glass structure is able to accommodate densification without the disruption its tetrahedral structure, by buckling and rotations of the network. Interestingly, SiO$_4$ tetrahedra remain almost rigid, giving the pressure-induced RUMs \cite{wells}. The slight increase of Si and O coordination numbers takes place in the 3--5 GPa, followed by their sharp increase for pressures exceeding 5 GPa \cite{p1,p2}. This has an important effect on network rigidity, because additional Si--O bonds that appear under pressure act as additional local constraints, increasing structural rigidity of the glass. We note that this effect is not related to the surface effects: because we use periodic boundary conditions, atoms at the boundary feel the same forces as atoms in the bulk.

\begin{figure}
\begin{center}
\rotatebox{-90}{\scalebox{0.5}{\includegraphics{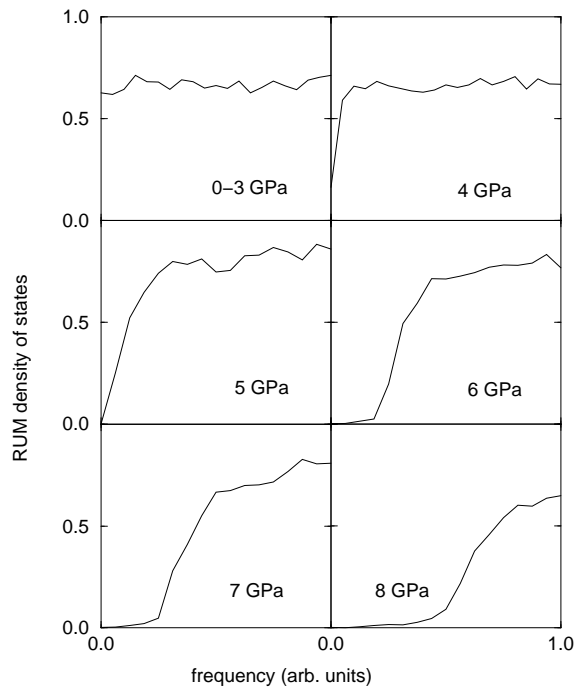}}}
\end{center}
\caption{The RUM density of states, calculated for SiO$_2$ glass compressed to pressures in the 0--8 GPa range.}
\label{fig5}
\end{figure}

We have quantified this effect by calculating the RUM densities of states for the densified structures \cite{p1,p2,p3,p4,p5}. The results are shown in Figure \ref{fig5}.

As follows from Figure \ref{fig5}, $g(\omega)$ at low $\omega$ decreases on increasing pressure, and comes to zero at 5 GPa, at which point $g(\omega)\propto\omega$. On further pressure increase, $g(\omega)$ flattens out at the origin, becoming $g(\omega)\propto\omega^2$. As discussed above, the presence of RUMs in our algorithm is witnessed by the non-zero value of $g(\omega)$ at $\omega=0$. For this reason, we define the pressure point at 5 GPa at which  $g(\omega)=0$ at $\omega=0$ as rigidity percolation point.

One can be more precise about the change of RUM flexibility at the rigidity percolation point. What actually happens is that RUMs do not completely disappear above 5 GPa, but the change is in the portion of $k$-space at which RUMs exist. Here we argue by analogy to the RUM flexibility of crystalline materials. The case where the RUM density of states $g(\omega)=const$ corresponds in a crystal to where the RUMs lie on two-dimensional surfaces of wave vectors in reciprocal space, and the case where $g(\omega)\propto\omega$ corresponds to one-dimensional lines of wave vectors. There is, of course, no natural analogue of the crystalline reciprocal space in a glass, but the real space picture is that the correlated RUM motions involve lines of tetrahedra in the former case and planes in the second case. Figure \ref{fig5} suggests that this is the situation on increasing pressure above 5 GPa. Thus for pressure above 5 GPa there is still some network flexibility, but at a lower level.

Pressure-induced rigidity percolation has a profound effect on relaxation in glass under high pressure and temperature. This will be discussed in the next section.

\section{Network flexibility and physical properties}

\subsection{Relaxation under high pressure and temperature: pressure window in SiO$_2$ and GeO$_2$ glasses}

\begin{figure}
\begin{center}
\rotatebox{-90}{\scalebox{0.55}{\includegraphics{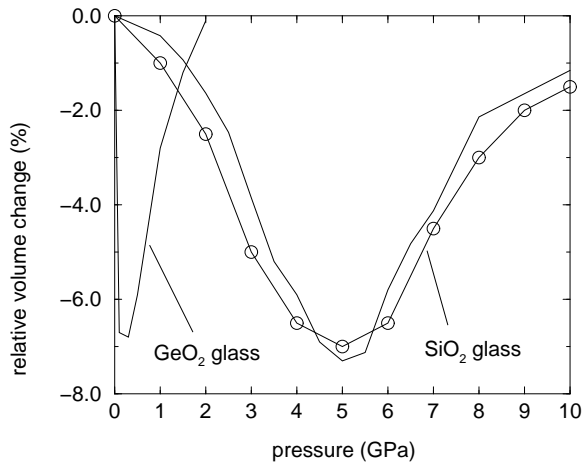}}}
\end{center}
\caption{Temperature-induced densification in the pressure window in SiO$_2$ and GeO$_2$ glasses. A circled line is the
experimental result; solid lines are the results from our MD simulation.}
\label{fig6}
\end{figure}

In the molecular dynamics simulations, we changed the pressure of the sample as a first stage, followed by a series of incremental changes in temperature \cite{p1,p2}. We define the temperature-induced in-situ densification as $\Delta V(P,T)=V(P,T)/V(P,T=300 K)-1)$, and plot $\Delta V$ in Figure \ref{fig6}. This Figure shows that at low and high pressures, $\Delta V$ is close to zero, but in a finite range of pressures around 5 GPa there is a significant decrease in $\Delta V$: $\Delta V$ becomes significantly nonzero and negative. Temperature-induced densification under pressure has, therefore, the form of a non-trivial {\it pressure window}.

The pressure window, predicted on the basis of molecular dynamics simulations \cite{p1}, was later confirmed in the in-situ high-pressure experiments (see Figure \ref{fig6}).

The pressure window in Figure \ref{fig6} can be explained by coupling of the structural rigidity to local relaxation events (LREs). At high pressure, a LRE is a localized large-scale ($\sim 1$ \AA) rebonding event that consists of breaking the existing bond(s), forming new one(s) and subsequent relaxation of the local structure \cite{p4}. A LREs is an irreversible process, which can be viewed as a ``quantum'' of structural change that a glass uses to adjust to changes of high pressure and temperature. An animation of a LRE is available in the electronic version of Reference \cite{p4}.

The pressure window is explained by noting that pressure has two competing effects on the dynamics of LREs. On one hand, pressure brings SiO$_4$ tetrahedra closer together, facilitating temperature-induced LREs by reducing their activation barriers and, therefore, increasing densification (recall that at high pressure, LREs are densification events). This is possible as long as large thermal motions, required by the LREs, are allowed to take place globally by the RUM flexibility of the glass. As we have seen in the previous section, this flexibility persists up to 5 GPa. On the other hand, there is a point at which pressure starts to reduce the RUM flexibility of the glass due to increased coordinations. This happens at 5 GPa. At this point, activation barriers for LREs increase markedly. This is because the structure looses the ability to sustain large-amplitude, low-energy RUMs required by LREs. At the rigidity percolation point, any large-amplitude motions require substantial energy cost related to breaking bonds (in contrast to very low energy cost when RUMs exist). As a result, the kinetics of LREs slows down markedly at the rigidity percolation point, resulting in densification decreasing after 5 GPa. The interplay of these two effects leads to the pressure window centered at 5 GPa. At this pressure, the effect is maximal because the tetrahedra are brought closest to each other to facilitate LREs and densification, and at the same time the structure is still flexible enough to allow for the LREs to take place globally.

We have predicted that pressure window also exists in GeO$_2$ glass (see Figure \ref{fig6}). We have found that in GeO$_2$ glass, increased coordinations appear in the structure at a much lower pressure, around 0.5 GPa. This is consistent with a recent study that employed a different interatomic potential \cite{mic}. The analysis of the RUM density of states showed that rigidity percolation set in almost simultaneously with the onset of increased coordinations. This gives rise to the temperature-induced densification in the pressure window located around 0.5-–1 GPa (see Fig. 1). The difference in location of the window with SiO$_2$ glass can be attributed to a smaller stiffness of tetrahedra in germania glass, which results in their deformation and rigidity percolation setting in at lower pressure and in a more narrow pressure interval than in SiO$_2$ glass.

The shape of the graph in Figure \ref{fig6} suggests the analogy with the ``reversibility window'' seen in chalcogenide glasses \cite{bool}. A large loss of irreversibility of the heat flow on cycling through the glass transition temperatures was seen in several systems. Interestingly, the reversibility window is located around the rigidity percolation point where $\langle r\rangle =2.4$ \cite{bool}. This is similar to our pressure window in \ref{fig6}, leading us to conclude that rigidity percolation can give interesting non-trivial effects in the relaxational behaviour of glasses as well as liquids.

\subsection{Crystallization and slow relaxation}

In the previous section, we have seen that the pressure windows shown in Fig. \ref{fig6} directly probe the changes of network rigidity of glasses under pressure. We now turn to other effects that can be understood on the basis of network rigidity. These are changes of crystallization temperature under pressure and slow logarithmic relaxation of pressurized glasses.

The crystallization temperature $T_c$ of SiO$_2$ glass under pressure has been measured \cite{ina}. It has been found that $T_c$ first decreases as pressure increases. This is followed by the sharp large increase of $T_c$ at about 7 GPa \cite{ina}. This behavior can be understood if we consider a LRE as an elementary relaxation in the path to crystallization. First, barriers to induce LREs decrease at low pressures, as the tetrahedra are brought closer to each other. Second, the kinetics of LRE is assisted by the RUM flexibility of the pressurized glass, which persists up to the pressure marking rigidity percolation. Therefore one expects initial decrease of the temperature needed to excite LREs and hence decrease of $T_c$. After rigidity percolates, energy barriers increase from the low values of RUM-type excitations to the considerably higher energies associated with the deformations of tetrahedra. Hence the temperature needed to induce LREs increases sharply after the rigidity percolation point, in good agreement with the experimental value.

We now consider the logarithmic relaxation of volume seen in SiO$_2$ and GeO$_2$ glasses under pressure \cite{bra}. In these experiments, it is found that for SiO$_2$ glass, logarithmic relaxation is only seen at pressures starting from about 7 GPa, while in GeO$_2$ glass it already sets in at about 2 GPa. The origin of this difference can be traced to the different response of network topology of the two glasses to pressure. We have suggested that the logarithmic relaxation can be adequately described by the dynamics of LREs \cite{jp-log}, in which the interaction between LRE-induced elastic fields plays a central role \cite{stress}. Structural rigidity is important here in the following sense. If no LREs are induced during pressurizing (which corresponds to the densification by mostly RUM-type distortions), no logarithmic relaxation is expected to take place. We now recall that SiO$_2$ glass densifies with the aid of RUM-type distortions coupled to a small number of LRE, up to 5 GPa, whereas in GeO$_2$ glass LREs are already induced at 0.5–-1 GPa. Hence in SiO$_2$ and GeO$_2$ glasses the logarithmic relaxation is expected to set in only after the corresponding points of rigidity percolation, in good agreement with the experimental observations \cite{bra}.

\subsection{Compressibility minimum}

Most materials get stiffer under pressure, due to the increased compaction of the constituent atoms. It is therefore interesting that some materials, including amorphous silica, actually get softer on compression; in the case of amorphous silica, there is a maximum in compressibility from experiment at a pressure of around 2 GPa \cite{bra}. This phenomenon remains unexplained.

We have recently explained this effect by discussing how network flexibility changes under pressure \cite{walker}. Note that here, pressures are small enough as not to disrupt the ideal tetrahedral network topology (see discussion above). Hence we consider the structural changes that take place within the ideal tetrahedral network. We have proposed that the compressibility minimum originates because at a certain pressure (around ambient pressure), RUMs result in the maximal atomic displacements (pressure- or temperature-induced). Note that this implies anharmonicity of RUMs. This proposal was based on the following general argument. At high pressure, the tetrahedra are compacted together, and the forces are such that the tetrahedra will deform and eventually bonds will be broken and reformed. These are high-energy processes, and hence the compressibility in this limit will be reduced. At negative pressure, leading to an expansion of the material, the network of tetrahedra will be stretched taut, and any changes in volume will require stretching of the covalent Si-O bonds. Once again, this is a relatively high-energy process, and the compressibility will be reduced. This implies the existence of the intermediate pressure, in which volume changes can be accommodated through buckling of the tetrahedral network with minimal energy cost (as discussed above, the necessary condition for such deformation is the presence of RUMs). In other words, at some intermediate pressure, a given amount of energy gives the RUM-type buckling of the structure with the maximal overall amplitude. This results in the compressibility minimum at that pressure.

A careful analysis of MD simulations revealed that the amplitudes of the rotational thermal motions of the SiO$_4$ tetrahedra are larger for intermediate pressures. This quantity acts as a proxy for the flexibility of the network, because the flexibility available for thermal fluctuations is also the flexibility available for accommodation for the buckling of the network under pressure \cite{walker}. This confirmed our argument about the origin of the compressibility minimum.

\subsection{Two-level systems}

In order to explain the anomalous heat capacity and thermal conductivity seen in glasses at low $\sim 1$ K temperature \cite{zel}, the phenomenological two-level systems (TLS) model has been proposed \cite{phi}. This model assumes the existence of additional, as compared to crystals, motions. These take place in localized atomic clusters which move in double-well potentials. At low temperature, the motion proceeds by quantum tunneling. In order to understand the microscopic details related to TLS, we have performed molecular dynamics simulations in SiO$_2$ glass \cite{ourjpcm,prl98}. These revealed reorientations of localized clusters of about 10 connected SiO$_4$ tetrahedra, and we suggested that these motions are candidate TLS in SiO$_2$ glass \cite{ourjpcm,prl98,prb-tls}.

We have recently revisited this issue to study whether TLS interact \cite{turl}. This is an important question, because, as assumed in the original model, TLS do not interact. However, it is conceivable that elastic fields from one TLS can affect the motion of atoms in another TLS. In this case, this presents us with a highly non-trivial problem where the low-energy states are not those associated with an isolated TLS, but are related to collective, due to interaction, states, with an important question of how to calculate energy states due to tunneling with strong long-range elastic interactions and whether we should view the additional motions as collective states from the outset. A recent work explored the model in which the collective low-energy states in glasses are formed by the softening of shear modulus on small length scales, without invoking tunneling as a required component of relaxation process \cite{turl0}.

\begin{figure}
\begin{center}
{\scalebox{0.42}{\includegraphics{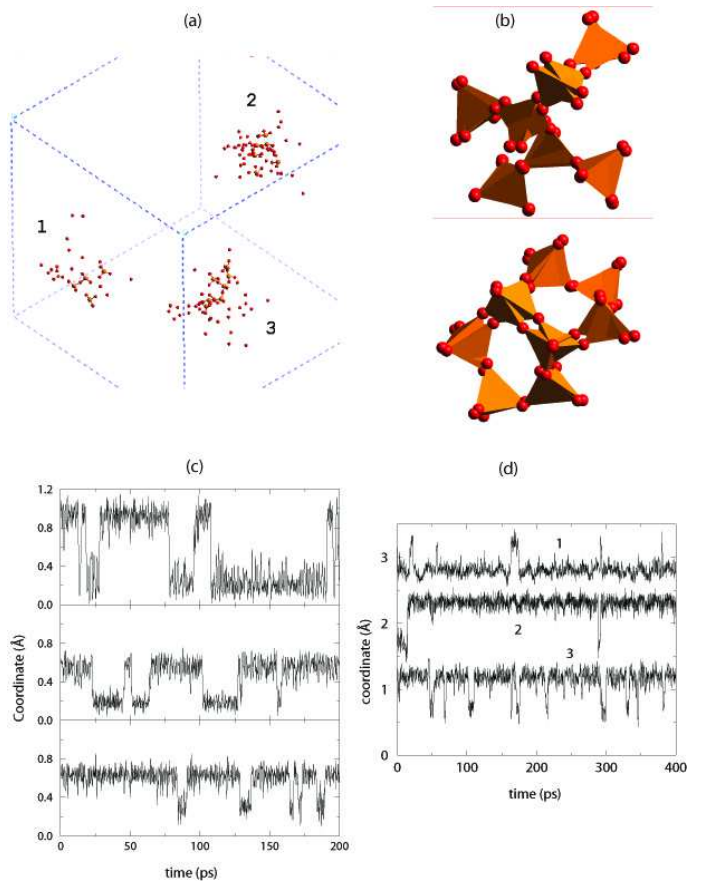}}}
\end{center}
\caption{(a) Three localized clusters undergoing large atomic jumps. Box size is about 58 \AA; (b) Examples of large-scale tetrahedral reorientations in each of the clusters in (a); (c) Examples of atomic trajectories within a cluster showing larger (top) and smaller (medium, bottom) magnitudes of atomic jumps; (d) Trajectories of atoms in clusters 1, 2 and 3 as shown in (a).}
\label{fig7}
\end{figure}

To understand the behaviour of low-energy states, we have performed extensive molecular dynamics simulations, in which we identified TLS as clusters of atoms which experience large jumps. In Figure \ref{fig7}a, we plot three TLS in the MD box. Figure \ref{fig7}b shows examples of motion in each TLS, seen as localized, nearly rigid, rotations and displacements of several connected SiO$_4$ tetrahedra. The values of atomic jumps are in the 0.3--0.8 \AA\ range, as illustrated in Figure \ref{fig7}c. Using the parameters of TLS from the simulation, in particular the hopping frequency in different clusters, we have estimated the average activation barrier for TLS of 178 K \cite{turl}. Using this value, the calculated tunnel splitting lies in the range of 0.01--0.8 K for different TLS. Note that the values of the activation barrier already include the effects of interaction between TLS (see below), corresponding to some effective barrier value.

The important effect is shown in Figure \ref{fig7}d, in which we observe that atomic jumps in different clusters take place at the same moments of time. It is seen that TLS 1 and 2 experience simultaneous jumps at 20 ps and 290 ps, TLS 1 and 3 at 170, 290, and 380 ps, and TLS 2 and 3 at 290 ps. All three TLS jump at 290 ps. On the scale from 0 to 1, the degree of correlation is in the range of 0.7--1, i.e. is significant \cite{turl}.

Hence, the simulations give TLS parameters not inconsistent with the original TLS model. At the same time, our simulations show that TLS strongly interact, and can not be considered independent as originally assumed. This presents a challenge for a future consistent theory of low-energy excitations in glasses. The low-energy hamiltonian of the problem may or may not contain tunneling terms, but it needs to contain interaction, according to our results.

Whatever the future theory holds for the low-energy dynamics in glasses, network flexibility of glass gives an important insight of what enables low-energy dynamics in the first place. Indeed, we have seen above that RUM flexibility ultimately determines whether or not a glass can support low-energy states. In crystals, RUMs are extended normal modes (in addition to trivial acoustic modes). Due to topological disorder in glass, RUMs can give rise to localized two-level systems which come in addition to usual phonons. These are the states we observed in Figure \ref{fig7}. The tetrahedral reorientations are RUM-like in that they do not involve distortion of SiO$_4$ units, and have relatively small activation barriers as a result. Were the system above the rigidity percolation point, low-energy TLS would be inhibited by large activation barriers related to bond breaking. Hence RUM flexibility is the {\it necessary condition} for existence of the localized low-energy excitations. This conclusion is independent of the way we eventually come to think about the low-energy states in glasses, i.e. independent local tunneling entities \cite{phi} or collective density excitations \cite{turl0}.

\section{Low-energy dynamics in oxide superconductors}

In the course of study of the origin of superconductivity in cuprates, the most controversial problem in condensed matter physics, several major shifts of opinion have taken place. Perhaps one of the most important of these is the recent realization that the electron-phonon interaction, largely discarded in the previous studies, actually plays an important role in superconductivity \cite{gunn,zhao,phil11}.

In discussing the low-energy vibrational states in cuprates, the Phillips-Thorpe constraint theory can used in the same way as discussed above, i.e. we can look for and identify the RUMs in the system (or in an important subsystem) and then study what effect they have on system properties. Recently, we have started to explore this idea \cite{philmag}.

\subsection{Low-energy states in the CuO$_2$ plane}

Lets consider the common structural unit of cuprate superconductors, the CuO$_2$ plane. It is accepted that persisting currents are formed in this plane. In order to discuss the ability of this system to support RUMs, one needs to identify rigid units, analogous to SiO$_4$ tetrahedra in silica. Cuprate superconductors are materials with mixed covalent and ionic bonding, hence, unlike in silica, their structures do not immediately offer the way to map them into the collection of rigid units that reflects the bonding type. A useful insight comes from the experimental and quantum-mechanical results that there is substantial covalency in Cu--O bonding in the Cu--O plane \cite{tajima}. Next, as in silicates, experiments point to the broken O bond-bending constraints \cite{phil2}. This allows us to consider the two-dimensional system of corner-shared rigid CuO$_4$ squares, loosely connected at corners (compare to a silicate, modeled in the RUM model as the system of rigid SiO$_4$ units, loosely connected at corners).

For this system, Maxwell counting gives the result that it is over-constrained, $N_c>N_f$. Indeed, each square has three degrees
of freedom in the plane, two translational and one rotational, giving $N_f=3$. There are two constraints per shared corner, or one constraint per corner per CuO$_4$ unit, or 4 constraints per unit in total, $N_c=4$. Hence Maxwell counting predicts that no RUM-type distortions should exist in CuO$_2$ plane. However, this approach does not take into account the important property of the system, its symmetry. It has been realized that symmetry can make certain constraints redundant, giving the flexibility of the system against RUMs. A detailed analysis of independent constraints shows that for the system shown in Figure 1, $N_f-N_c=4$ \cite{dove-trac}. Hence, in addition to two uniform translations and one rotation of the whole object, there exists one non-trivial floppy mode for this system. The nature of this mode can be studied in detail using the RUM analysis discussed in the previous sections. For a plane in the perovskite structure, this method finds the RUM with $k$ point at the zone boundary, which corresponds to rotations and displacements of units as shown in Figure 2 \cite{dove-trac}. We can readily extrapolate this result to the two-dimensional analogue of the perovskite structure, the system of connected CuO$_4$ squares, and find that this system has the optic RUM as is shown in Figure 2. Note the anisotropy of the RUM-induced field of atomic displacements, discussed below.

From the topological point of view, one can therefore view the two-dimensional system of connected squares in Figure 2 as an
interesting borderline and, perhaps, unique, case for its ability to support low-energy vibrations. It is neither under-constrained as to give many RUMs and hence small correlation length, nor over-constrained as to inhibit RUMs altogether. The balance of the degrees of freedom and the number of constraints, together with the system's symmetry, give the result that only one single RUM (at the single k point) is present. As a result, the displacement pattern propagates in this two-dimensional system into the entire plane (see Figure 2), giving the infinite correlation length for the coordinate correlation function \cite{philmag}. It should be noted that this is true only in the model in which the CuO$_4$ units are infinitely rigid. In practice, there is always a finite distortion of the units, leading to a finite correlation length which, however, exceeds the size of the unit.

We note that the frequency of the RUM mode is zero only in the RUM model, and in real cuprates, it is defined by the inter-unit and other next-order interactions, as well as by the effects of steric hindrance on the CuO$_2$ plane. During the distortion in Figure 2, one pair of O atoms comes closer to the out-of-plane cations directly above the square centre (for example, in La$_2$CuO$_4$, these are La/Sr ions; in YBa$_2$Cu$_3$O$_7$, these are Y and Ba ions), whereas the remaining pair comes closer to the cations above the centres of neighbouring squares. The energy of these interactions sets the scale of the RUM frequency. A large body of data exists on phonons in cuprates. In the context of superconductivity, most of the recent discussion has concentrated on the behaviour of higher-energy breathing and half-breathing modes. On the other hand, the behaviour of the low-frequency RUM has not been discussed, apart from earlier neutron scattering experiments \cite{pint-rev,reich}. In these experiments, the RUM was identified as the zone boundary optic phonon with frequency in the 3--3.6 THz range in the tetragonal phase of Nd$_2$CuO$_4$, Pr$_2$CuO$_4$, La$_2$CuO$_4$, and 6.3 THz in YBCO.

We also note that in this discussion, we do not consider low-frequency out-of-plane tilting motion of the octahedra. These vibrations vary in different cuprates due to different out-of-plane environments, and can result in material-specific phase transitions at material-dependent temperatures (below we discuss the effect of the associated distortion of CuO$_2$ plane on the RUM). Here, we discuss the RUM that is {\it generic} to the CuO$_2$ plane.

\subsection{Negative thermal expansion}

It is interesting to note that negative thermal expansion, decrease of volume (or some of the system's linear dimensions) on temperature increase, is related to the presence of RUMs in a system \cite{r4}. As discussed above, the RUM distortion pulls the structure onto itself, as the distance between the centres of the units decreases (see Figure 2). If this effect exceeds usual thermal expansion, the net effect can be volume reduction on temperature increase.

This effect is seen in the in several cuprates as the anomalous change of the unit cell parameters in CuO$_2$ plane. In YBCO, for example, deviation from the linear decrease of the parameters of CuO$_2$ plane is seen around superconducting temperature $T_{\rm c}$, followed by the negative thermal expansion at lower temperature \cite{youprb}. We can therefore interpret these effects as the manifestation of existence of the RUM in the CuO$_2$ plane.

\subsection{Relevance for superconductivity}

After its identification in the early neutron scattering experiments \cite{pint-rev,reich}, the RUM in Figure 2 has not been discussed for its relevance for superconductivity. Instead, the discussion focused on the higher-energy breathing and half-breathing modes in the CuO$_2$ plane.

By construction, the RUM is the lowest vibrational energy state in the CuO$_2$ plane (apart from trivial acoustic modes), because it does not involve bond-stretching and bond-bending distortions that come at high energy cost. Any other non-trivial in-plane mode has necessarily higher frequency because it involves bending and stretching bonds (for example, the energy of the bond-stretching breathing mode is 18--20 THz \cite{pint-rev}, several times larger than the RUM frequency). Being the low-frequency vibration, the RUM couples strongly to charge carriers. In BSCCO, for example, the spectral function shows strong coupling in the 12--27 meV range \cite{gonnelli} ($\approx$3--6 THz), which is in the range of the RUM frequency. Of course, it is important to calculate the electron-phonon coupling constant directly from the phonon-induced deformation of the electronic structure. We are currently exploring this problem.

The important observation is that the anisotropic pattern of the RUM displacement is consistent with the d-wave symmetry of the order parameter \cite{philmag}. It follows from Figure 2 that the electron-phonon interaction is strongest along the Cu--O bonds, and is weakest along the directions that run diagonal to Cu--O bonds. This, together with strong coupling of the RUM to charge carriers, gives large anisotropic gap with d-wave symmetry, as is seen experimentally in cuprates.

This is an important insight, because the observed d-wave symmetry in cuprates is often taken as evidence against the phonon mechanism of superconductivity and is used to support other theories. This approach overlooks the fact that if a structure has anisotropy and two-dimensional flexible subsystems (such as CuO$_2$ planes in cuprates), phonons can, in fact, give d-wave symmetry of the order parameter.

The relevance of the RUM for superconductivity is consistent with the correlation between $T_{\rm c}$ and the structural state of the CuO$_2$ plane. If the RUM plays a role, one expects that $T_{\rm c}$ should be maximized in the system in which the RUM is favoured most. Experimentally, for different cuprates at a given doping level, $T_{\rm c}$ is maximal in structures in which the CuO$_2$ plane is flat and square \cite{mourach}. In our picture, this behaviour can be rationalized by noting that the RUM is most favoured when the CuO$_2$ plane is as close to being flat and square as possible. On the other hand, distortion of the plane results in suppression of the RUM (see Figure 2). The distortion can be, for example, plane buckling with large out-of-plane strains, although these can be partially compensated by slightly different Cu--O distances \cite{bianconi}. Another type of distortion can be rhombic distortion of the plane, during which the RUM serves as a soft mode. This results in suppression of the RUM due to lower symmetry of the orthorhombic phase relative to the tetragonal one.

Finally, it is interesting to note that in cuprates, the charge in the CuO$_2$ plane is ordered in stripes that run either along
Cu-Cu-Cu or Cu-O-Cu bonds and fluctuate in the transverse direction \cite{mourach}. This pattern is consistent with the anisotropic field of the RUM displacements (see Figure 2). This hints to a possible relationship between the RUM-induced field and the dynamic stripe order.

\section{Summary}

In summary, we have discussed several important relaxation processes that involve low-frequency modes. These modes can be determined from viewing a solid as a mechanical network with a certain degree of rigidity. We reviewed the low-frequency modes in several important oxide materials, in SiO$_2$ glass at ambient and high pressure as well as in cuprate superconductors. We discussed how the RUMs affect the properties of these materials. In SiO$_2$ glass, the ability to support RUMs governs its relaxation in the wide range of pressures and temperatures, giving rise to the non-trivial ``pressure window''. It also affects other properties, including crystallization, slow relaxation and compressibility minimum. At ambient pressure and low temperature, RUM flexibility is related to the large-scale localized atomic motions. Whether these motions are interpreted as independent two-level systems or collective density excitations, the RUM flexibility determines whether and to what extent low-energy excitations can exist in a given glass structure. Finally, we discussed the RUM in the CuO$_2$ plane in cuprate superconductors, and its relevance for the d-wave symmetry of the order parameter and other properties.

We are pleased to acknowledge important and stimulating discussions with Professors J. C. Phillips and V. V. Brazhkin and support from EPSRC.


\end{document}